\documentstyle[epsf,aps]{revtex}

\begin{document}

\setlength{\textwidth}{7.0in}
\setlength{\textheight}{8.5in}
\setlength{\evensidemargin}{0in}
\setlength{\oddsidemargin}{0in}
\setlength{\topmargin}{-0.5in}

%\twocolumn

%\report{}

%\draft{}
 
\newcounter{univ_counter}
\setcounter{univ_counter} {0}
\addtocounter{univ_counter} {1} 
\edef\FSU{$^{\arabic{univ_counter}}$ } 

\addtocounter{univ_counter} {1} 
\edef\ASU{$^{\arabic{univ_counter}}$ } 

\addtocounter{univ_counter} {1} 
\edef\UCLA{$^{\arabic{univ_counter}}$ } 

\addtocounter{univ_counter} {1} 
\edef\CMU{$^{\arabic{univ_counter}}$ } 

\addtocounter{univ_counter} {1} 
\edef\CUA{$^{\arabic{univ_counter}}$ } 

\addtocounter{univ_counter} {1} 
\edef\SACLAY{$^{\arabic{univ_counter}}$ } 

\addtocounter{univ_counter} {1} 
\edef\CNU{$^{\arabic{univ_counter}}$ } 

\addtocounter{univ_counter} {1} 
\edef\UCONN{$^{\arabic{univ_counter}}$ } 

\addtocounter{univ_counter} {1} 
\edef\DUKE{$^{\arabic{univ_counter}}$ } 

\addtocounter{univ_counter} {1} 
\edef\EDINBURGH{$^{\arabic{univ_counter}}$ } 

\addtocounter{univ_counter} {1} 
\edef\FIU{$^{\arabic{univ_counter}}$ } 

\addtocounter{univ_counter} {1} 
\edef\GWU{$^{\arabic{univ_counter}}$ } 

\addtocounter{univ_counter} {1} 
\edef\ORSAY{$^{\arabic{univ_counter}}$ } 

\addtocounter{univ_counter} {1} 
\edef\ITEP{$^{\arabic{univ_counter}}$ } 

\addtocounter{univ_counter} {1} 
\edef\INFNGE{$^{\arabic{univ_counter}}$ } 

\addtocounter{univ_counter} {1} 
\edef\INFNFR{$^{\arabic{univ_counter}}$ } 

\addtocounter{univ_counter} {1} 
\edef\JMU{$^{\arabic{univ_counter}}$ } 

\addtocounter{univ_counter} {1} 
\edef\KYUNGPOOK{$^{\arabic{univ_counter}}$ } 

\addtocounter{univ_counter} {1} 
\edef\MIT{$^{\arabic{univ_counter}}$ } 

\addtocounter{univ_counter} {1} 
\edef\UMASS{$^{\arabic{univ_counter}}$ } 

\addtocounter{univ_counter} {1} 
\edef\UNH{$^{\arabic{univ_counter}}$ } 

\addtocounter{univ_counter} {1} 
\edef\NSU{$^{\arabic{univ_counter}}$ } 

\addtocounter{univ_counter} {1} 
\edef\OHIOU{$^{\arabic{univ_counter}}$ } 

\addtocounter{univ_counter} {1} 
\edef\ODU{$^{\arabic{univ_counter}}$ } 

\addtocounter{univ_counter} {1} 
\edef\PITT{$^{\arabic{univ_counter}}$ } 

\addtocounter{univ_counter} {1} 
\edef\RPI{$^{\arabic{univ_counter}}$ } 

\addtocounter{univ_counter} {1} 
\edef\RICE{$^{\arabic{univ_counter}}$ } 

\addtocounter{univ_counter} {1} 
\edef\URICH{$^{\arabic{univ_counter}}$ } 

\addtocounter{univ_counter} {1} 
\edef\SCAROLINA{$^{\arabic{univ_counter}}$ } 

\addtocounter{univ_counter} {1} 
\edef\UTEP{$^{\arabic{univ_counter}}$ } 

\addtocounter{univ_counter} {1} 
\edef\JLAB{$^{\arabic{univ_counter}}$ } 

\addtocounter{univ_counter} {1} 
\edef\VT{$^{\arabic{univ_counter}}$ } 

%\addtocounter{univ_counter} {1} 
%\edef\VSU{$^{\arabic{univ_counter}}$ } 

\addtocounter{univ_counter} {1} 
\edef\VIRGINIA{$^{\arabic{univ_counter}}$ } 

\addtocounter{univ_counter} {1} 
\edef\WM{$^{\arabic{univ_counter}}$ } 

\addtocounter{univ_counter} {1} 
\edef\YEREVAN{$^{\arabic{univ_counter}}$ } 

\title{{\large Electroproduction of the $\Lambda(1520)$ hyperon }}
%%%%%%%%%%%%%%%%%%%% authors %%%%%%%%% 
 \author{ 
S.P.~Barrow,\FSU\
L.~Dennis,\FSU\
S.B.~McAleer,\FSU\
G.~Adams,\RPI\
M.J.~Amaryan,\YEREVAN\
E.~Anciant,\SACLAY\
M.~Anghinolfi,\INFNGE\
%D.S.~Armstrong,\WM\
B.~Asavapibhop,\UMASS\
G.~Asryan,\YEREVAN\
G.~Audit,\SACLAY
% ...
\\
\begin{center}
(The CLAS Collaboration)
\end{center}
\vspace{-0.15in}
} 
\address{\FSU Florida State University, Tallahasee, Florida 32306}
\address{\ASU Arizona State University, Tempe, Arizona 85287-1504}
\address{\UCLA University of California at Los Angeles, Los Angeles, California  90095-1547}
\address{\CMU Carnegie Mellon University, Pittsburgh, Pennslyvania 15213}
\address{\CUA Catholic University of America, Washington, D.C. 20064}
\address{\SACLAY CEA  Saclay, DAPNIA/SPhN, F91191 Gif-Sur-Yvette Cedex, France}
\address{\CNU Christopher Newport University, Newport News, Virginia 23606}
\address{\UCONN University of Connecticut, Storrs, Connecticut 06269}
\address{\DUKE Duke University, Durham, North Carolina 27708-0305}
\address{\EDINBURGH Edinburgh University, Edinburgh EH9 3JZ, United Kingdom}
\address{\FIU Florida International University, Miami, Florida 33199}
\address{\GWU The George Washington University, Washington, DC 20052}
\address{\ORSAY Institut de Physique Nucleaire ORSAY, Orsay, France}
\address{\ITEP Institute of Theoretical and Experimental Physics, Moscow, 117259, Russia}
\address{\INFNGE Istituto Nazionale di Fisica Nucleare, Sezione di Genova and Dipartimento di Fisica dell'Universit\`{a}, 16146 Genova, Italy }
\address{\INFNFR Istituto Nazionale di Fisica Nucleare, Laboratori Nazionali di Frascati, Frascati, Italy}
\address{\JMU James Madison University, Harrisonburg, Virginia 22807}
\address{\KYUNGPOOK Kyungpook National University, Taegu 702-701, South Korea}
\address{\MIT Massachusetts Institute of Technology, Cambridge, Massachusetts  02139-4307}
\address{\UMASS University of Massachusetts, Amherst, Massachusetts  01003}
\address{\UNH University of New Hampshire, Durham, New Hampshire 03824-3568}
\address{\NSU Norfolk State University, Norfolk, Virginia 23504}
\address{\OHIOU Ohio University, Athens, Ohio  45701}
\address{\ODU Old Dominion University, Norfolk, Virginia 23529}
\address{\PITT University of Pittsburgh, Pittsburgh, Pennslyvania 15260}
\address{\RPI Rensselaer Polytechnic Institute, Troy, New York 12180-3590}
\address{\RICE Rice University, Houston, Texas 77005-1892}
\address{\URICH University of Richmond, Richmond, Virginia 23173}
\address{\SCAROLINA University of South Carolina, Columbia, South Carolina 29208}
\address{\UTEP University of Texas at El Paso, El Paso, Texas 79968}
\address{\JLAB Thomas Jefferson National Accelerator Laboratory, Newport News, Virginia 23606}
\address{\VT Virginia Polytechnic Institute and State University, Blacksburg, Virginia   24061-0435}
%\address{\VSU Virginia State University, Petersburg,Virginia 23806}
\address{\VIRGINIA University of Virginia, Charlottesville, Virginia 22901}
\address{\WM College of Willliam and Mary, Williamsburg, Virginia 23187-8795}
\address{\YEREVAN Yerevan Physics Institute, 375036 Yerevan, Armenia}

\date{\today}
\maketitle

\begin{abstract}
\begin{center}
{\large Abstract}
\end{center}
The reaction $ep \rightarrow e'K^+\Lambda(1520)$ 
with $\Lambda(1520) \rightarrow p'K^-$  was studied at electron
beam energies of 4.05, 4.25, and 4.46 GeV,
using the CLAS detector at the Thomas Jefferson
National Accelerator Facility.
The cos$\theta_{K^+}$, $\phi_{K^+}$, $Q^2$, 
and $W$ dependencies of $\Lambda$(1520) electroproduction are presented
for the kinematic region 0.9 $<$ $Q^2$ $<$ 2.4 GeV$^2$ and 1.95 $<$ $W$ $<$ 2.65 GeV.
Also, the $Q^2$ dependence of the $\Lambda$(1520) decay angular distribution is presented for
the first time.
The cos$\theta_{K^+}$ angular distributions suggest {\it t}-channel
diagrams dominate the production process.
Fits to the $\Lambda$(1520) {\it t}-channel helicity frame 
decay angular distributions indicate the 
$m_{z}=\pm\frac{1}{2}$ parentage accounts for about 60\%
of the total yield, 
which suggests this reaction has a significant contribution from {\it t}-channel processes
with either   K$^+$ exchange or longitudinal coupling to an exchanged K$^*$.
The $Q^2$ dependence of the $\Lambda$(1520) production cross section
is the same as that observed for $\Lambda$(1116)
photo- and electroproduction.

\end{abstract}

\twocolumn
%\onecolumn

\begin{center}
{\large I. Introduction}
\end{center}

Many important discoveries in nuclear and particle physics,
such as CP violation, were initially observed in 
hadrons containing strange quarks.
Current studies of strange quark phenomena are
motivated by such issues as the importance of the strange quark-antiquark 
sea within nucleons and the predicted abundance of strange 
quarks within the quark-gluon plasma.
The strange quark also introduces a new degree of freedom
into the nuclear medium and thus provides a unique new look at
conventional nuclear physics through the study of hypernuclei.
Studies of strange baryon electroproduction have
been noticeably missing.

During the 1970's there were two published measurements 
of $\Lambda$(1520) photoproduction [1-2], as well as
one electroproduction measurement \cite{aze75}, and since then
there have been no further published studies of these reactions.
The construction of the Continuous Electron Beam Accelerator Facility (CEBAF) at the
Thomas Jefferson National Accelerator Facility (Jefferson Lab) with its high duty cycle beam
and modern detectors has provided a unique new opportunity to resume the study
of strange baryon
photo- and electroproduction.
This paper reports the first measurement of $\Lambda$(1520) electroproduction
that used the CEBAF Large Acceptance Spectrometer (CLAS) in Hall B of  Jefferson Lab.

One of the photoproduction measurements \cite{bar80} used beam energies
from 2.8 to 4.8 GeV (total
center-of-mass energy $W$ from 2.5 to 3.1 GeV), and 
reports an exponential {\it t}-dependence 
dominated by {\it t}-channel exchange of the $K^{*}$(892) meson,
and not the lighter $K$(494) meson.
A measurement \cite{boy71} at higher  photon energies 
also yields an exponential {\it t}-dependence to the cross section.
The lone electroproduction measurement \cite{aze75}  concludes  the 
variation of the cross section with the 
virtual photon invariant mass $Q^2$ 
from 0.1 to 0.5 GeV$^2$ is consistent
with a simple vector meson dominance model.
These groundbreaking measurements were difficult due to the
limited data sample sizes.

There are several motivations for further study of $\Lambda$(1520) electroproduction.
Theoretical models \cite{wjc92,dav96} for the electroproduction of the lighter
$\Lambda$(1116) predict
large contributions from the longitudinal virtual photon cross section.
Similar behavior in $\Lambda$(1520) electroproduction could result
in an enhanced $K$(494) {\it t}-channel exchange relative to the
photoproduction result \cite{bar80}.
Such a possibility emphasizes the importance of measuring the 
relationship between $Q^2$ and $K$(494) exchange. 
Recently, Capstick and
Roberts \cite{cap88} predicted the existence of several nonstrange 
N$^*$ resonances with significant ($\sim 5\%$) branching ratios into the 
$\Lambda$(1520) + $K^+$ decay channel.
Extending the $Q^2$ range of the $\Lambda$(1520) electroproduction measurement allows
an additional examination of resonance contributions to $\Lambda$(1520) production.
Furthermore, $\Lambda$(1520) electroproduction from a hydrogen
target necessitates the creation of a strange quark-antiquark pair. 
Although the kinematic regime studied in this work is typically
associated with hadronic degrees of freedom, 
it is nonetheless important  to search for  any evidence of quark degrees of freedom
in strange baryon production.

In addition to new insight into  $\Lambda$(1520) 
production, the current results represent a significant new step
in the study of hyperon production phenomenology.
For the first time, it will be possible to make quantitative 
comparisons of the $Q^2$ dependencies of the $\Lambda$(1520), 
$\Lambda$(1116),  and $\Sigma$(1193) cross sections. Hopefully this
information will stimulate theoretical efforts to
model $\Lambda$(1520) electroproduction, especially since currently no
published theory papers discuss it apart from Ref. \cite{cap88}.   

In the current experiment, the CLAS detector was used to
study the decay
angular distribution of the electroproduced $\Lambda$(1520), as well as the
dependencies on
$W$, $Q^2$, and the center-of-mass
angles $\phi_{K^+}$ and $\cos \theta_{K^+}$.
The data
span  the region of 
$Q^2$ from 0.9 to 2.4 GeV$^2$, and $W$ up to 2.65 GeV. 
The large acceptance and high multiplicity capabilities
of CLAS make it possible to study $\Lambda$(1520) production over
this  wide kinematic region. Details
about the experiment and data analysis are discussed in Section II. 
Section III presents the 
results from the current analysis. In Section IV the results
are summarized and compared with previous measurements and
theoretical interpretations of 
$\Lambda$(1116) and $\Sigma$(1193) production.

\begin{center}
{\large II. Experiment}
\end{center}

\begin{figure}
\epsfysize203mm
\centerline{\epsfbox{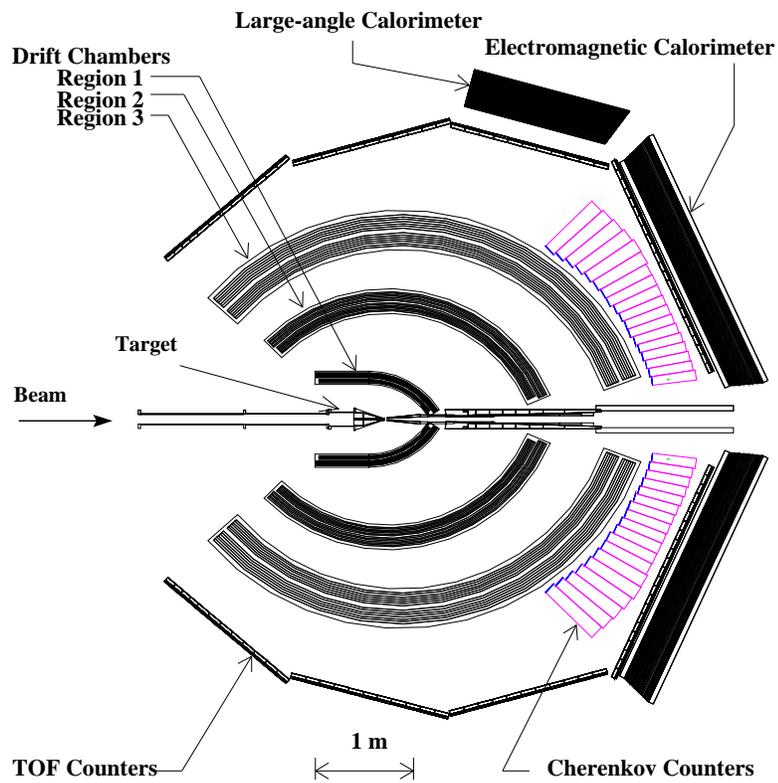}}
\caption{\small{A cross sectional view of the CLAS detector through
two opposing sectors. The direction of the $e^-$ beam is from
left to right. }}
\label{fig:clas}
\end{figure}

The CLAS detector \cite{clas_nim}, shown
schematically in  Fig.~\ref{fig:clas}, 
is a six sector toroidal magnetic spectrometer.
This design deflects charged particles toward or
away from the beam line while leaving the particle's
azimuthal angle unchanged.
Six wedge-shaped sectors surround the beam line.
The three drift chamber \cite{dc1,dc2} regions per sector 
are used to measure
the momentum vector and charge of all tracks.
Each sector also contains 48 scintillator paddles \cite{sc}
to determine the event start time and the 
hadron masses, Cherenkov detectors \cite{cc}
to distinguish between electrons and negatively charged pions,
and calorimeters \cite{ec,lac} to identify neutral particles, 
as well as to assist with the e$^-$/$\pi^-$ separation. 

The data presented in this paper are the accumulated total
for experiment E89-043 
from more  than 42 days of data taking during 
the 1998 and 1999 E1 run periods.
These E1 run periods used electron beam
energies of 4.05, 4.25, and 4.46 GeV,
incident on a 
liquid hydrogen target.
The electron beam current was typically 4.5 nA, 
which yielded a nominal
luminosity of about 6$\times$10$^{33}$ cm$^{-2}$s$^{-1}$,
and a total integrated luminosity of 5$\times$10$^{39}$ cm$^{-2}$.
Seventeen CLAS experiments ran concurrently during these runs,
which was accomplished by the use of
an inclusive electron trigger \cite{trig92}.
Roughly 2 billion events were recorded 
but less than 0.5\% of them correspond to reconstructed $\Lambda$(1520)
electroproduction events.

In order to study decay angular distributions of the $\Lambda$(1520),
it is necessary to detect the scattered electron, the
$K^{+}$, and one of the decay fragments from a binary decay
channel of the $\Lambda$(1520). The final state
$e^-$-$K^{+}$-$p$,  with an undetected $K^{-}$
reconstructed using missing mass techniques,
is best suited for study with CLAS 
for these E1 run periods. During these runs the 
toroidal  magnetic field  was oriented such that positively charged 
particles were bent away from the beam pipe.
The $\Lambda$(1520) $\rightarrow$
$pK^{-}$ decay channel
accounts for  22.5\% \cite{PDG} of its total width. 
The main issues in identifying this decay mode
of electroproduced $\Lambda$(1520)'s are briefly discussed below,
and further discussions of these topics
are presented in Refs. \cite{costy2k} and \cite{clasnote}.

\begin{center}
{\it a. Particle identification}
\end{center}

Reconstruction of CLAS data starts with
the identification of the electron.
Electron candidates 
create a shower in the calorimeter consistent with the momentum
of the track as defined by the drift chambers,
and also generate a signal in the Cherenkov
detector.
Once an electron is identified, its path length
and the TDC information from the time-of-flight
scintillation paddle it traverses are used to determine the 
event start time. This information is then used to determine the flight time 
for the hadron tracks, which,  combined with the
reconstructed hadron momentum, determines the mass
for charged tracks.

\begin{figure}[htb] 
\epsfysize133mm
\centerline{\epsfbox{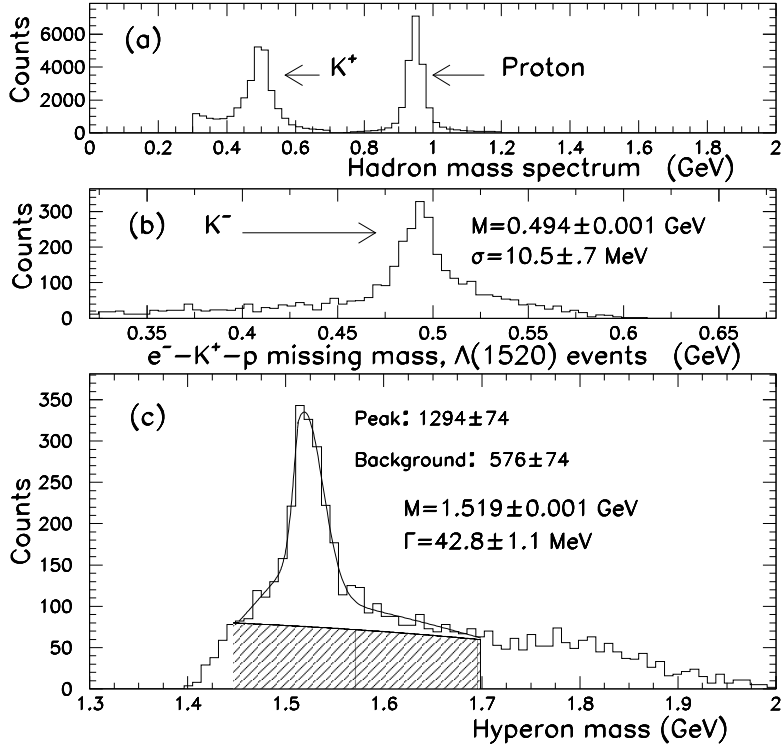}}
\caption{\small{(a) The hadron mass spectrum
for events that contain a proton track and  a $K^+$
candidate.
 (b) The $K^-$ missing mass spectrum for events in which
the e$^-$-$K^+$ missing mass is consistent with the $\Lambda$(1520)
mass. (c) The hyperon mass spectrum for the e$^-$-$K^+$-$K^-$-$p$
final state. A cut on the $K^-$ missing mass from 0.455 to 0.530 GeV was used to generate
this hyperon spectrum. }}
\label{fig:hadrons}
\end{figure}

Figure~\ref{fig:hadrons}(a) shows
the hadron mass spectrum
for events that contain a proton track as well as  a $K^+$
candidate. 
Proton and   $K^+$ tracks are selected by
appropriate cuts on this spectrum. The $K^+$
mass cut is a function of the momentum of the track
to compensate for the diminished mass resolution
as the speed of the $K^+$  approaches
the speed of light.

Monte Carlo simulations of CLAS indicate that events in which the $K^{+}$ decays prior to traversing
the time-of-flight scintillators are the largest contribution to the background in the
$K^{-}$  missing mass spectrum shown in Figure~\ref{fig:hadrons}(b). These $K^{+}$ decays are properly
modeled and accounted for in our Monte Carlo acceptance calculations.

There also exists a fairly significant monotonically decreasing background
under the $K^+$ peak in Fig.~\ref{fig:hadrons}(a). This is due to
high momentum $\pi^+$ tracks that have a large uncertainty in their reconstructed
mass. These misidentified tracks do not introduce a significant source of background
to the $\Lambda$(1520) data set, since the events containing
these tracks seldom generate a missing mass consistent
with the $K^{-}$ mass cut. The contribution from events containing
misidentified $\pi^{+}$ tracks  in Fig.~\ref{fig:hadrons}(b) is 
less than 1\% of the total yield.

The $\Lambda$(1520) centroid and width  plotted in Fig.~\ref{fig:hadrons}(c) 
are based
on a fit to a Gaussian with a radiated tail for the $\Lambda$(1520) 
peak, and a fourth-order
polynomial parameterization of  the background
for the hyperon mass region from 1.44 to 1.70 GeV. 
Given the nominal full width at half the maximum
(FWHM) of 15.6 MeV \cite{PDG} for the $\Lambda$(1520) mass,
the measured FWHM of 42.8 MeV shown in Fig.~\ref{fig:hadrons}(c), 
indicates the intrinsic FWHM resolution
of CLAS for this reaction is about 39 MeV. Therefore the width of the $\Lambda$(1520)
peak in Fig.~\ref{fig:hadrons}(c) is dominated by the experimental resolution.
The parameterization of the background is indicated by the shaded region 
in Fig.~\ref{fig:hadrons}(c).

The resolution  for reconstructing 
$Q^2$ and $W$ is about 1\%. The resolution for
hadronic scattering angles 
varies from $\sim$ 0.2 to $\sim$ 1.2 degrees, depending
on whether the scattering angle is a function of  one or both
of the reconstructed hadrons. For example, the reconstructed electron, $K^+$,
and proton  are  needed to calculate the $\Lambda$(1520)
helicity frame decay angles, whereas only the electron and  $K^+$
are used to calculate the center-of-mass angle $\theta_{K^+}$.

\begin{figure}[htb] 
\epsfysize143mm
\centerline{\epsfbox{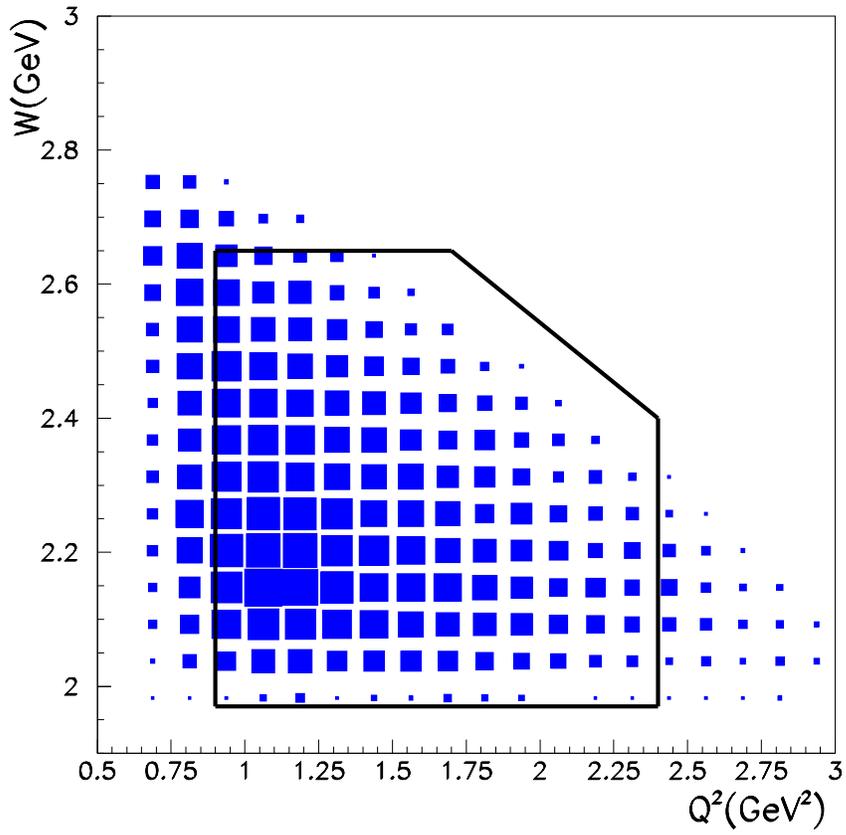}}
\caption{\small{The region of $Q^2$ versus $W$ discussed in this paper
is bounded by the solid lines. The data included in this figure
are the same $\Lambda$(1520) events presented in the other
figures of this paper.}}
\label{fig:grid}
\end{figure}

The region of $Q^2$ versus $W$ included in this paper 
is shown in Fig.~\ref{fig:grid}. The lower limit of 
$Q^2$ $=$ 0.9 GeV$^2$ was chosen in order to have a common cutoff
for the data taken with the 4.05  and 4.25 GeV electron beam energies. 
The upper $Q^2$ cutoff at 2.4 GeV$^2$ is due to limited
statistics for higher $Q^2$.

\begin{center}
{\it b. Backgrounds}
\end{center}

Reactions that produce other hyperons,
such as the $\Lambda(1405)$, $\Sigma$(1480), and $\Lambda$(1600),
account for the majority of the background under the
$\Lambda$(1520) peak, but
the relative contributions from the
individual processes are currently unknown.
A complete listing of the hyperons whose mass and width have
some overlap with the $\Lambda$(1520) peak 
is presented in Ref. \cite{PDG}.

Another possible source of background in Fig.~\ref{fig:hadrons}(c)
is from the $K^+$-$K^-$ decay of  
$\phi$(1020) meson production.
However, simulations \cite{costy99}
of the acceptance of CLAS for $\phi$(1020) and $\Lambda$(1520) electroproduction
indicate there is little overlap between these two
processes. The $\Lambda$(1520) reaction is by far the dominant
one, and the contamination due to the $\phi$(1020)
meson is at the level of 1-2\%.

The $\Lambda$(1520) background was studied as a 
function of $Q^2$, $W$, cos$\theta_{K^+}$,
and $\phi_{K^+}$. 
The only significant dependency
in the background was for cos$\theta_{K^+}$ (and correspondingly, {\it t}),
in which the background ranged from  25\% of the total
yield for cos$\theta_{K^+}$$\sim$1, to a 45\% contribution for cos$\theta_{K^+}$
close to $-1$. 
The methods used to parameterize
the background in the helicity frame decay angular distributions are discussed in
Section III({\it{b}}).

\begin{center}
{\it c. Cross sections}
\end{center}

Cross sections were calculated using the following definition
of the virtual photon flux factor:

\begin{equation}
\Gamma = \frac{\alpha}{4\pi}\frac{W}{E^2M^2}(W^2-M^2)\frac{1}{Q^2}\frac{1}{1-\varepsilon}.
\end{equation}

\noindent Here $\alpha$ is the fine structure constant,
and $M$ and $E$ are the proton mass and electron beam energy, respectively.
The transverse polarization of the virtual photon, $\varepsilon$,
has the standard definition:

\begin{equation}
\varepsilon = \left(1+2\frac{|{\vec Q}|^2}{Q^2}\tan^2\frac{\theta_e}{2}\right)^{-1},
\end{equation}

\noindent and $\theta_e$ is the polar scattering angle of the electron in the laboratory
frame.

The $\Lambda$(1520) cross sections shown in this paper are
derived from acceptance
corrected, normalized yields in the hyperon mass region from
1.492 to 1.555 GeV. These yields are scaled upward to 
compensate for the tails
of the $\Lambda$(1520) distribution that lie outside this interval.
The acceptance of the CLAS detector was derived from a Monte Carlo
simulation that folded the $K^+$ decay 
into the geometric acceptance. 
The cross sections are corrected for experimental
dead time,  track reconstruction efficiency, and
contributions from the walls of the target cell.
Radiative corrections
were calculated following the Mo and Tsai approach \cite{mo69}.
The combined systematic error of the cross sections 
from these  corrections is about 9\%, and is mainly due
to the geometric acceptance corrections.
The yields are also scaled downward,
typically by 25-30\%,  to correct for the presumed incoherent
background
under the $\Lambda$(1520) peak.
The parameterization of the
hyperon background introduces an additional systematic
uncertainty in the $\Lambda$(1520) cross sections of approximately  10\%.

\begin{center}
{\large III. Results \\}
\end{center}

Details about the cos$\theta_{K^+}$ and {\it t}-distributions are the first results
presented. 
Section III({\it b}) presents
the main results of this paper, the decay angular
distributions of the $\Lambda$(1520).
Section III({\it c}) shows plots
related to the virtual photon cross sections and the scattered electron degrees of freedom.
 
\begin{center}
{\it a. cos$\theta_{K^+}$ and t-distributions}
\end{center}

\begin{figure}[htb] 
\epsfysize143mm
\centerline{\epsfbox{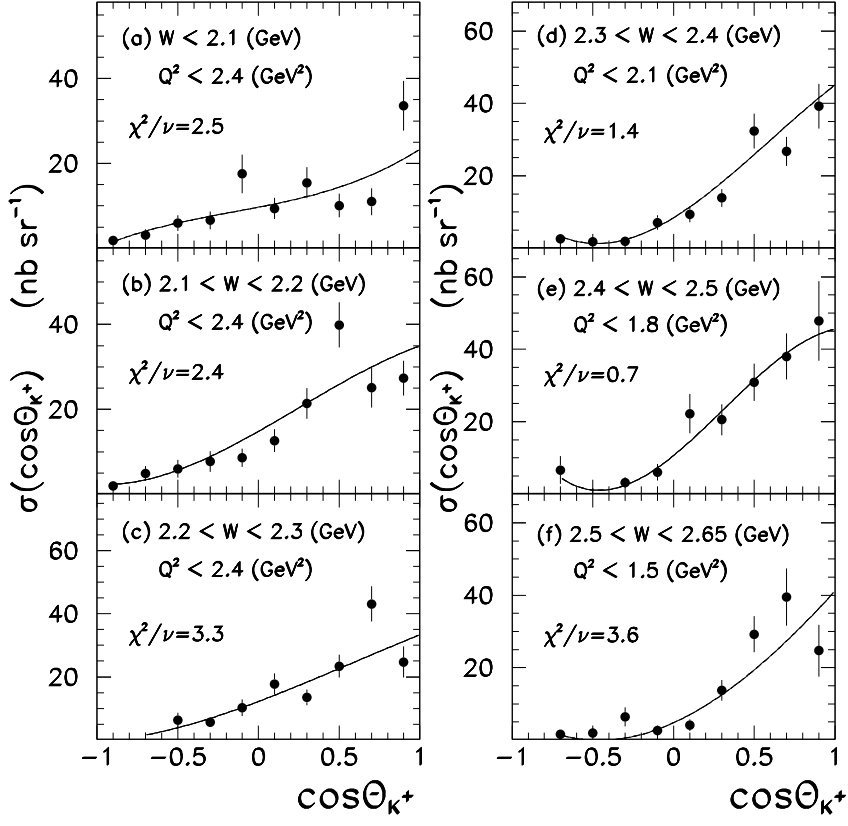}}
\caption{\small{The cos$\theta_{K^+}$ differential
cross section distributions for six regions of $W$.
The error bars are statistical uncertainties only.
The solid lines represent Legendre polynomial fits that are described in the text.
The lower limit $Q^2$ $=$ 0.9 GeV$^2$ is used for all six distributions.}}
\label{fig:w_dep_theta}
\end{figure}

The dependence of the cross section  on cos$\theta_{K^+}$ for six regions
of $W$ is shown in Fig.~\ref{fig:w_dep_theta}. Throughout this paper
cos$\theta_{K^+}$ is defined to be
the center-of-mass angle subtended by the outgoing $K^+$
and the direction of the incident virtual photon
in the rest frame of the virtual photon and
the target proton.
The curves plotted in these
figures are the results of fits to the first four Legendre
polynomials, $\sum_{i=0}^{i=3} a_{i}P_{i}$, and the
normalized fitted coefficients are summarized in Table 1.
These fits provide a simple
parameterization of the variation of the 
cos$\theta_{K^+}$ distributions with $W$. 
The coefficient $a_{0}$ slowly increases in strength
as $W$ approaches threshold. In addition, there is clearly some $W$ 
dependence to $a_{2}$, the coefficient
of the l$=$2 Legendre polynomial, which is larger
at higher $W$ than near threshold, and
the fit at the highest $W$ bin only qualitatively reproduces
the data. It is possible
both of these effects are due
to enhanced $K^*$(892) exchange at higher $W$.

\begin{figure}[htb] 
\epsfysize143mm
\centerline{\epsfbox{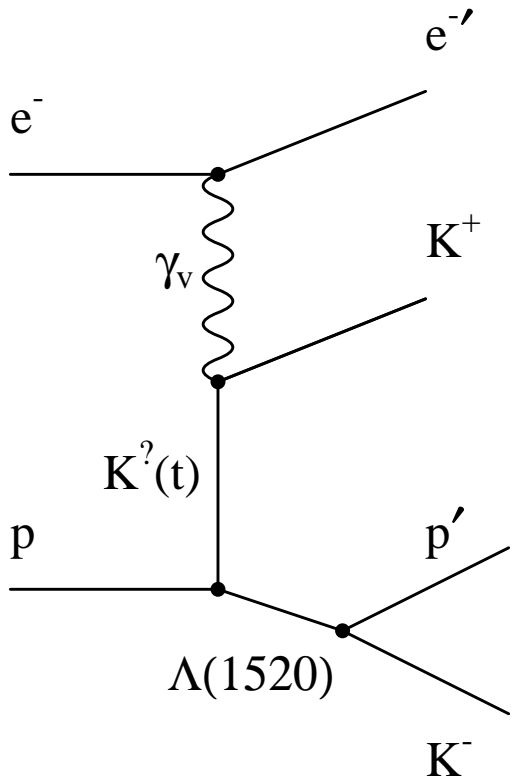}}
\caption{\small{The generic {\it t}-channel process for $\Lambda$(1520) 
electroproduction, for events in which the $\Lambda$(1520) decays into the
$p + K^-$ final state. The exchanged kaon is denoted as
$K^?$ since there are several kaons that could be exchanged.
The four-vector for this exchanged meson is  $t$, as is
indicated in this figure.}}
\label{fig:t_def}
\end{figure}

\begin{figure}[htb] 
\epsfysize133mm
\centerline{\epsfbox{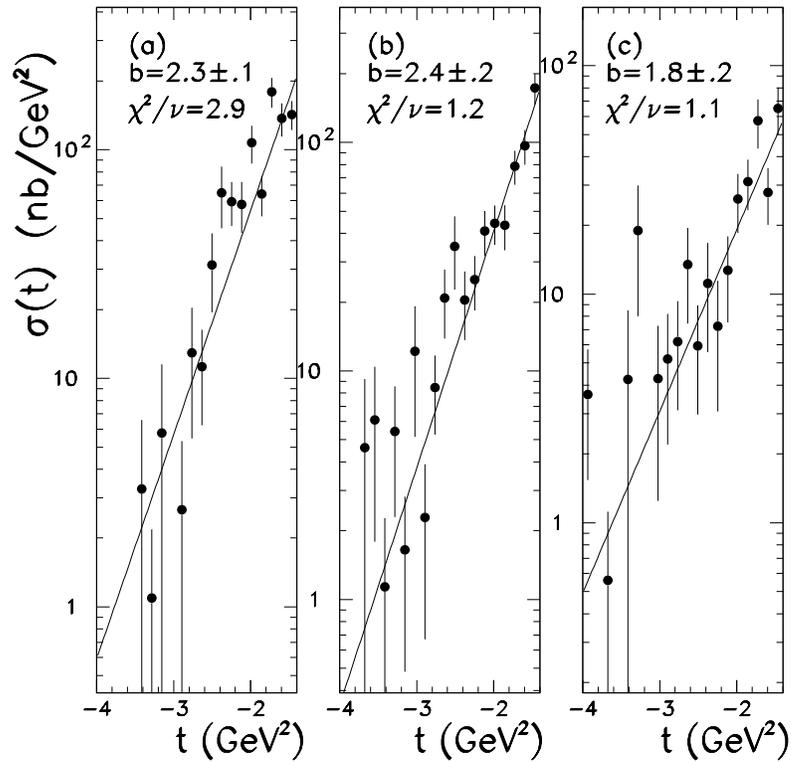}}
\caption{\small{The {\it t}-distributions for three regions of $W$: (a)
1.95 $<$ $W$ $<$ 2.21 GeV, (b) 2.21 $<$ $W$ $<$ 2.43 GeV, and (c) 2.43 $<$ $W$ $<$ 2.65 GeV.
The fitted value of the exponent of the exponential, $b$, 
is indicated in each plot,
along with the reduced $\chi^2$ of the fit. The uncertainties
indicated for the values of $b$ are due to statistical uncertainties only.
The parameterization of the background under
the $\Lambda$(1520) peak contributes an additional
uncertainty of $\sim$.07 to $b$.}}
\label{fig:t_dist}
\end{figure}

If the distributions are instead plotted versus 
{\it t}, the squared magnitude of the exchanged meson 4-vector
shown in Fig.~\ref{fig:t_def},
the data are fairly well parameterized by
the exponential $e^{bt}$ for 
{\it t} from $-3.7$ to $-1.4$ GeV$^2$, as is shown in Fig.~\ref{fig:t_dist}.
No significant $W$ dependence to $b$ is observed.
Our electroproduction value for $b$
of 2.1 $\pm$ 0.3 GeV$^{-2}$ indicates a reduction of the 
interaction region \cite{costy2k} relative 
to a  photoproduction measurement \cite{bar80}, 
which reports an $e^{(6.0)t}$ behavior 
for {\it t} from $-0.65$ to $-0.25$ GeV$^2$.

\begin{figure}[htb] 
\epsfysize23mm
\centerline{\epsfbox{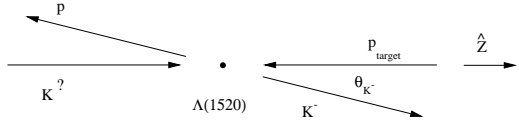}}
\caption{\small{The definition of the {\it t}-channel
helicity frame angle $\theta_{K^-}$. }}
\label{fig:helicity_def}
\end{figure}

Since there is no evidence for cross section
strength at large $\theta_{K^+}$ angles for any $W$, there does not appear to be
appreciable {\it s}-channel resonance contributions.
Instead, both the cos$\theta_{K^+}$ and the {\it t}-distributions
are consistent with the behavior expected for {\it t}-channel dominance.
Therefore, the $\Lambda$(1520)
decay angular distributions will be presented in the {\it t}-channel
helicity frame. The {\it t}-channel diagram for this reaction
is shown in Fig.~\ref{fig:t_def}.
Following the convention of Ref. \cite{bar80},
the {\it t}-channel
helicity frame {\it z}-axis is defined to be antiparallel to the
direction of the incident proton in the $\Lambda$(1520) rest frame,
as is illustrated in Fig.~\ref{fig:helicity_def}, 
and the {\it y}-axis is normal to the hyperon production plane.

\begin{center}
{\it b. Helicity frame distributions}
\end{center}

\begin{figure}[htb] 
\epsfysize143mm
\centerline{\epsfbox{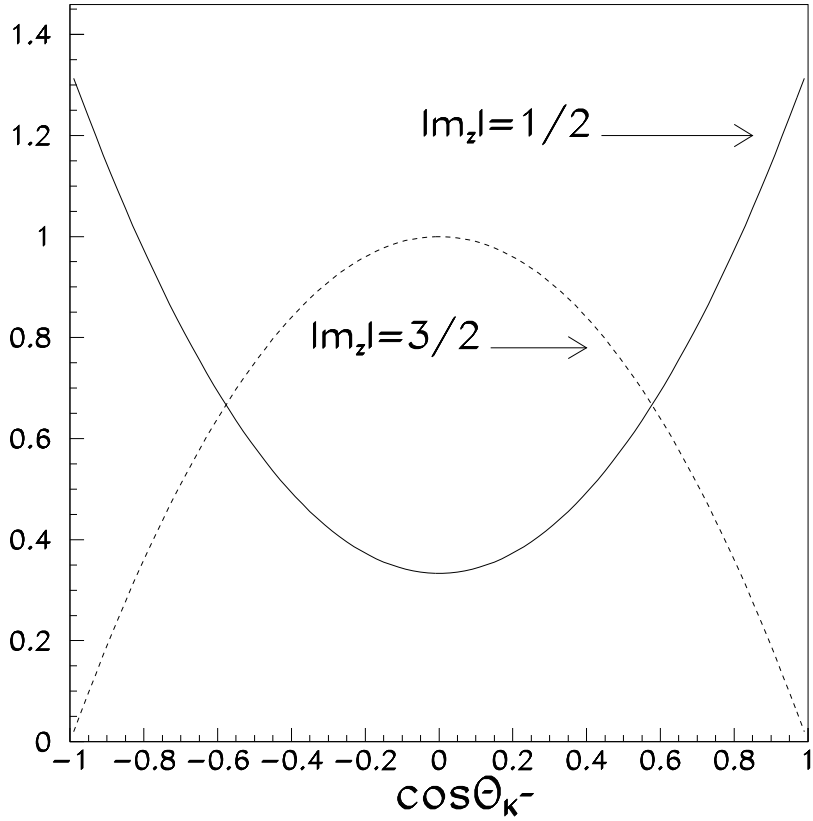}}
\caption{\small{The expected {\it t}-channel helicity frame
decay angular distributions if the $\Lambda$(1520) hyperons
were produced exclusively in the m$_z$=$\pm {1 \over 2}$ spin projections
(solid line) or m$_z$=$\pm {3 \over 2}$ (dashed line) projections.}}
\label{fig:form}
\end{figure}

The $\Lambda$(1520) is a J$^{\pi}$ $=$ $\frac{3}{2}^-$ baryon,
and its $p$-$K^-$ decay is a parity
conserving strong decay mode.
A straightforward application of Clebsch-Gordon geometry
demonstrates that for  an $m_{z}=\pm\frac{3}{2}$ projection the decay 
is characterized by a sin$^{2}\theta_{K^-}$
distribution, while an $m_{z}=\pm\frac{1}{2}$ projection
has a $\frac{1}{3}$+cos$^{2}\theta_{K^{-}}$
distribution. These distributions are illustrated in Fig.~\ref{fig:form}.

\begin{figure}[htb] 
\epsfysize143mm
\centerline{\epsfbox{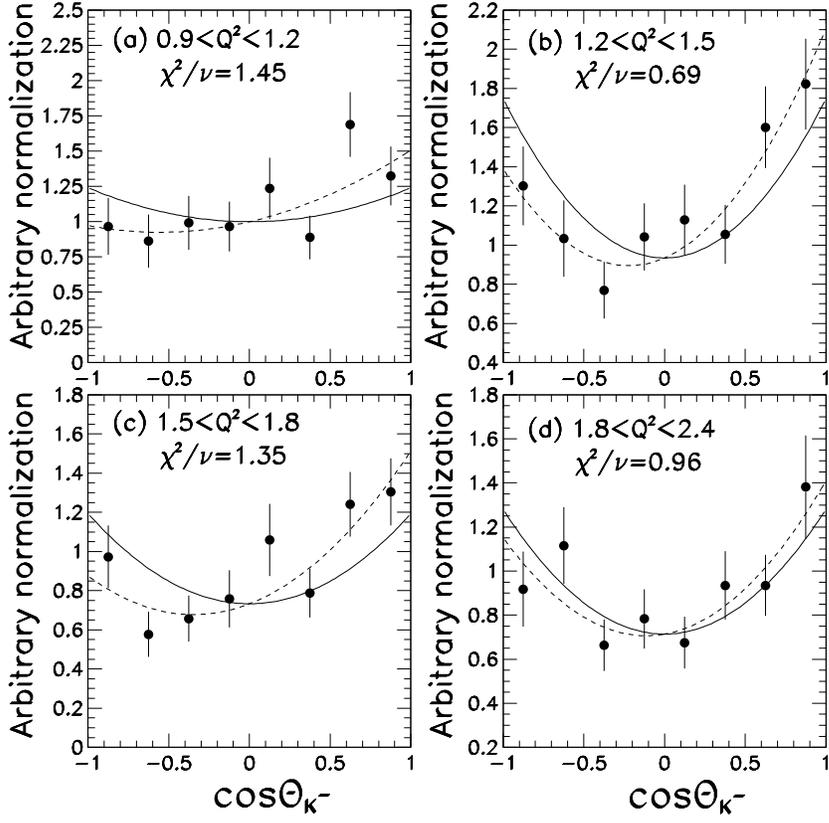}}
\caption{\small {The $\Lambda$(1520) cos$\theta_{K^-}$ decay angular
distribution for four regions of $Q^2$. These distributions
are averaged over the region of $W$ from threshold to 2.43 GeV.
The error bars are statistical uncertainties only. The plotted curves
are explained in the text.}}
\label{fig:q_dep}
\end{figure}

\begin{figure}[htb] 
\epsfysize143mm
\centerline{\epsfbox{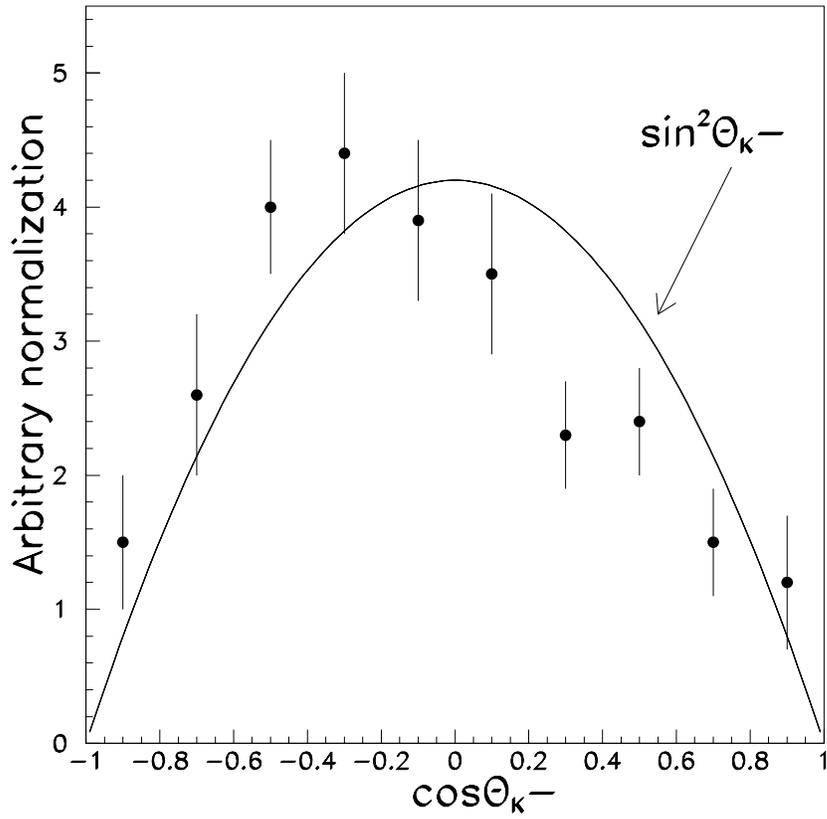}}
\caption{\small {The $\Lambda$(1520) photoproduction
decay angular distribution published previously
in Ref. [1]. The sin$^2\theta_{K^-}$ curve included
with this data is the expected distribution if the
$\Lambda$(1520) decay is entirely due to the m$_z$=$\pm {3 \over 2}$
spin projection, and is not a fit to the data.}}
\label{fig:bar_fig}
\end{figure}

The {\it t}-channel helicity frame cos$\theta_{K^{-}}$ 
decay angular distributions for four regions
of $Q^2$ are shown in Fig.~\ref{fig:q_dep}.
Also shown in this figure are plotted curves that are described below.
The analogous distribution for the photoproduction result \cite{bar80} is shown
in Fig.~\ref{fig:bar_fig}. It is clear from a visual inspection of these
two figures that the current results represent a significant
departure from what was measured in Ref. \cite{bar80}.
The photoproduction angular distribution possesses
a greatly enhanced $m_{z}=\pm\frac{3}{2}$ parentage relative to
the electroproduction results presented here. 
All four of the distributions shown in Fig.~\ref{fig:q_dep}
demonstrate a large
$\frac{1}{3}$+cos$^{2}\theta_{K^{-}}$ contribution, which indicates
the electroproduced $\Lambda$(1520) hyperons 
are primarily populating the $m_{z}=\pm\frac{1}{2}$ spin projection.

If $\Lambda$(1520) electroproduction proceeds exclusively through {\it t}-channel
exchange of a spinless kaon, the 
$\Lambda$(1520) spin projection is always  $m_{z}=\pm\frac{1}{2}$,
and the ratio  of the $m_{z}=\pm\frac{3}{2}$
to $m_{z}=\pm\frac{1}{2}$ 
populations is zero.
On the other hand, if the reaction  proceeds exclusively through 
the transverse exchange of a J$=$1 $K^*$ vector meson, the ratio of the $m_{z}=\pm\frac{3}{2}$
to $m_{z}=\pm\frac{1}{2}$ 
spin projections, if solely determined
by Clebsch-Gordon coefficients, is 3 to 1.
Therefore the  electroproduction
distributions  shown in  Fig.~\ref{fig:q_dep} could be evidence for a roughly
equal mixture of $K^*$(892) and $K$(494) contributions.
In contrast, the photoproduction result \cite{bar80} suggests that reaction
proceeds almost exclusively through transverse $K^*$(892) exchange.

Each dashed line plotted in 
Fig.~\ref{fig:q_dep} is the result of a fit to the two
$\Lambda$(1520) spin projection distributions 
with an additional cos$\theta_{K^{-}}$ term:

\begin{eqnarray}
f(\theta_{K^{-}})=\alpha(\frac{1}{3}+\cos^{2}\theta_{K^{-}})+\beta\sin^{2}\theta_{K^-}+\gamma\cos\theta_{K^{-}}.
\end{eqnarray}

\begin{figure}[htb] 
\epsfysize143mm
\centerline{\epsfbox{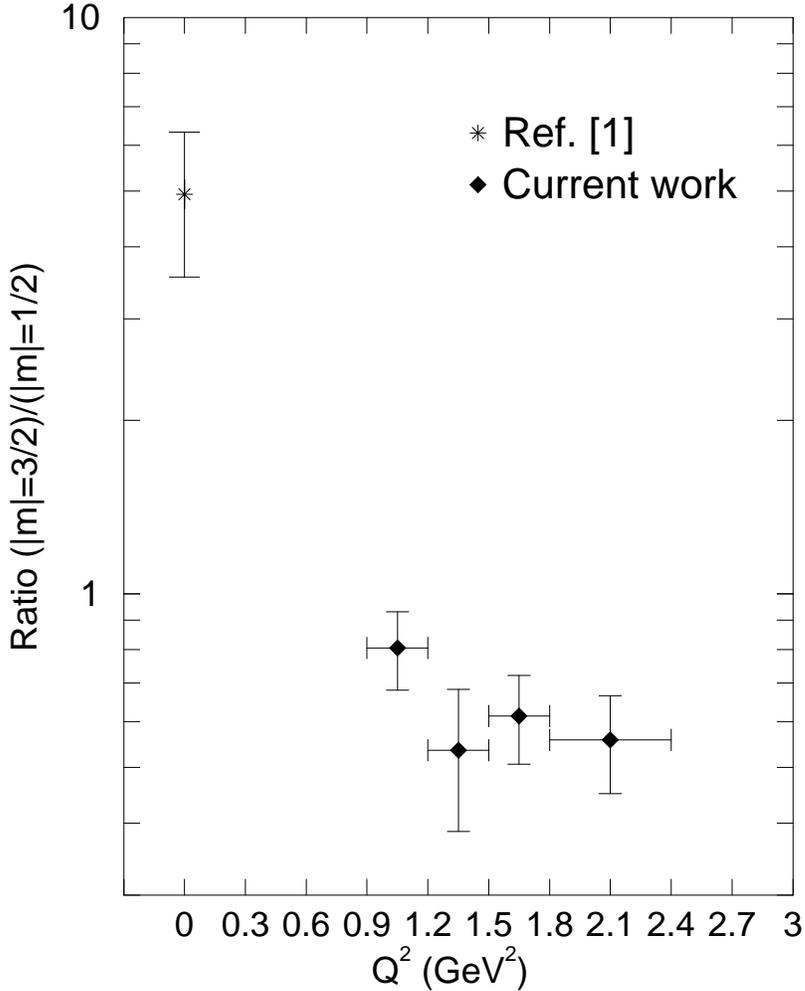}}
\caption{\small{The ratios $\frac{(|m|=3/2)}{(|m|=1/2)}$ 
of the spin projection populations,
based on  the ratios $\frac{\beta}{\alpha}$ of the fitted parameters in
Equation 3, for each region of $Q^2$.
The point at $Q^2$ = 0 is derived  from Fig. 3
of Ref. [1]. The vertical error bars are derived from the uncertainties
of the fitted  coefficients $\alpha$ and $\beta$. The horizontal error bars denote the
averaging intervals. }}
\label{fig:new_ratio}
\end{figure}

\noindent These are the only fits that were used to analyze these distributions.
The solid lines in Fig.~\ref{fig:q_dep} are the contribution to each fit from just the two
$\Lambda$(1520) decay angular distribution terms.  The spin 
projection parentages are  derived from the
ratios of the fitted parameters $\alpha$ and $\beta$.
Figure ~\ref{fig:new_ratio} plots the spin projection ratios
for these four regions of $Q^2$, along with the result
from the photoproduction measurement \cite{bar80}.
The electroproduction ratios are summarized in Table 2.

Roughly two-thirds of the known hyperons \cite{PDG}  that
overlap the $\Lambda$(1520)
have spin J=$\frac{1}{2}$.
Coherently combining the angular distributions
from a J $=$ $\frac{1}{2}$ background 
with the  J $=$ $\frac{3}{2}$ $\Lambda$(1520) decay yields several
interference terms possessing cos$\theta_{K^{-}}$ terms
raised to odd powers. 
The  cos$\theta_{K^{-}}$ contributions to the decay angular distributions
could therefore be evidence of  J $=$ $\frac{1}{2}$ background hyperons.
The photoproduction decay angular distribution
shown in Fig.~\ref{fig:bar_fig} also indicates the existence
of a weak cos$\theta_{K^{-}}$ contribution.

The J=$\frac{1}{2}$  hyperons possess
flat helicity frame decay angular distributions.
If a flat angular distribution is fit
to the two  $\Lambda$(1520)
spin projection distributions,  the result will be an
even mixture of the two projections, since
($\frac{1}{3}$ + cos$^{2}\theta_{K^{-}}$) + (sin$^{2}\theta_{K^-}$)=constant.
If some of the background under the $\Lambda$(1520) peak is
due to J=$\frac{1}{2}$ hyperons, it  will make
equal contributions to the two spin projections, and
artificially shift the measured 
spin projection ratio closer to one. This is true
regardless of whether the actual ratio for the  $\Lambda$(1520) spin
projections is greater than or less than 1. 
Additional information about the physical processes
that contribute to the background  is needed to estimate
this effect more quantitatively.

\begin{figure}[htb] 
\epsfysize143mm
\centerline{\epsfbox{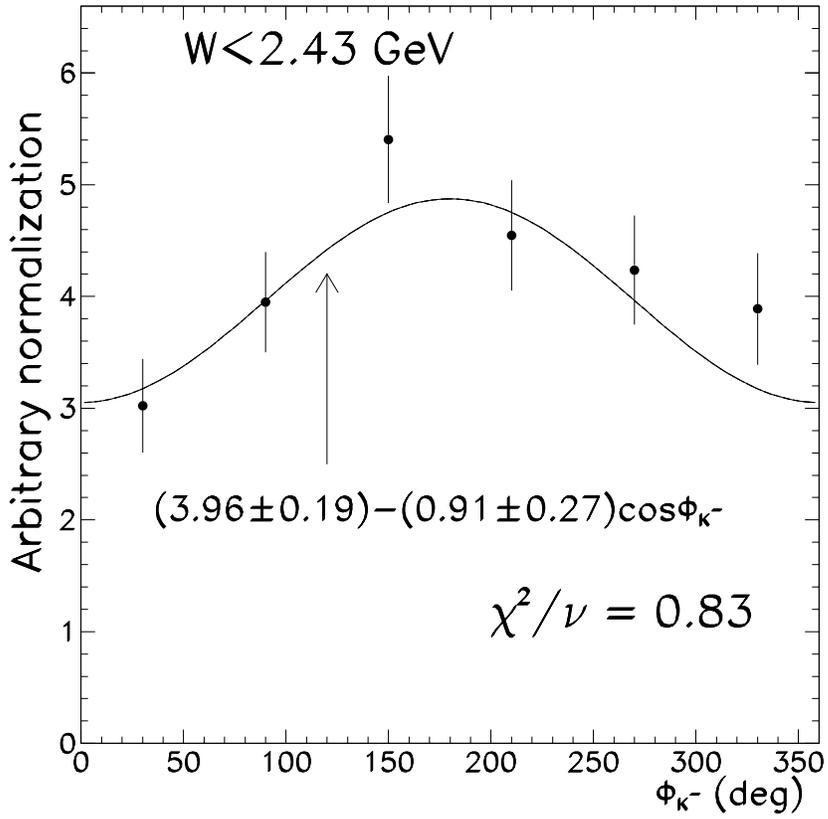}}
\caption{\small{The $\phi_{K^-}$ decay angular distribution for $W$ $<$ 2.43 GeV.
Also plotted is the result of a fit of the form A+B$*$cos$\phi_{K^-}$.
The error bars are statistical uncertainties only. }}
\label{fig:phi_all}
\end{figure}

The {\it t}-channel helicity frame 
$\phi_{K^{-}}$ decay angular distribution for $W$ $<$ 2.43 GeV, 
summed over the entire range
of $\theta_{K^-}$, is shown  in Fig.~\ref{fig:phi_all}.
The fit plotted in Fig.~\ref{fig:phi_all} includes a cos$\phi_{K^-}$ term,
and indicates this term makes an important contribution.
An isolated J=$\frac{3}{2}$ resonance does not possess 
a cos$\phi_{K^-}$ dependence, therefore, 
as was the case for the cos$\theta_{K^{-}}$ term added to the fits
in Fig.~\ref{fig:q_dep}, this cos$\phi_{K^-}$ dependence could also
be due to interference effects with other hyperons.

\begin{center}
{\it c. $\phi_{K^+}$, W, and Q$^2$ distributions}
\end{center}

The dependence on $\phi_{K^+}$, the angle between the hadron and lepton scattering planes,
is sensitive to the relative contributions
of the longitudinal and transverse components of the virtual photon.
This is illustrated in the following
decomposition of  the center-of-mass cross section, 

\begin{eqnarray}
\sigma(W,Q^2,\theta_{K^+},\phi_{K^+})\sim\sigma_T+\varepsilon\sigma_L+\\
\nonumber \varepsilon\sigma_{TT}\cos2\phi_{K^+}+\sqrt{\frac{\varepsilon(\varepsilon+1)}{2}}\sigma_{LT}\cos\phi_{K^+}.
\end{eqnarray}

\begin{figure}[htb] 
\epsfysize143mm
\centerline{\epsfbox{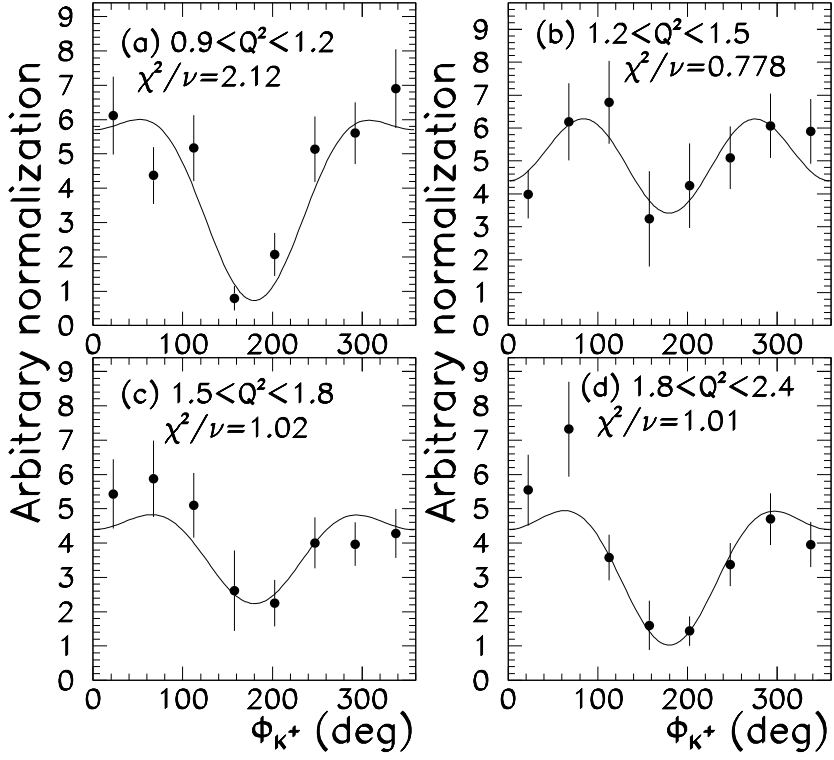}}
\caption{\small{The $\phi_{K^+}$ distributions for the same four regions of
kinematics shown in Fig.~9. The plotted curves are the results
of fits of the form A+B$\ast$cos2$\phi_{K^+}$+C$\ast$cos$\phi_{K^+}$.
The plotted error bars are statistical uncertainties only.
The results of those fits, with the constant term normalized to one,
are summarized in Table 3.}}
\label{fig:volker}
\end{figure}

\noindent The $\sigma_{LT}$ term is only an indirect
measurement of the relative contributions of the longitudinal and
transverse cross sections. If it makes a large contribution, we
expect that both the longitudinal and transverse couplings of the virtual photon
are significant.
Figure ~\ref{fig:volker}
plots the $\phi_{K^+}$ distributions for the same four regions of Q$^2$
shown in Fig.~\ref{fig:q_dep}.
The range of $\varepsilon$, the transverse polarization
of the virtual photon,  for the data presented here is
from 0.3 to 0.7 with a nominal value
of $\sim$ 0.5. The fits shown in Fig.~\ref{fig:volker} are summarized in Table 3.
All four fits suggest cos$\phi_{K^+}$ contributions,
indicating contributions from both
the longitudinal and transverse virtual photon spin
projections. However, there is a larger Q$^2$ dependence to 
this term than to the ratios
of the spin projections shown in Fig.~\ref{fig:new_ratio}.
This demonstrates that the virtual photon L-T interference does not have
a direct correspondence with the L-T decomposition
of the helicity frame.

\begin{figure}[htb] 
\epsfysize133mm
\centerline{\epsfbox{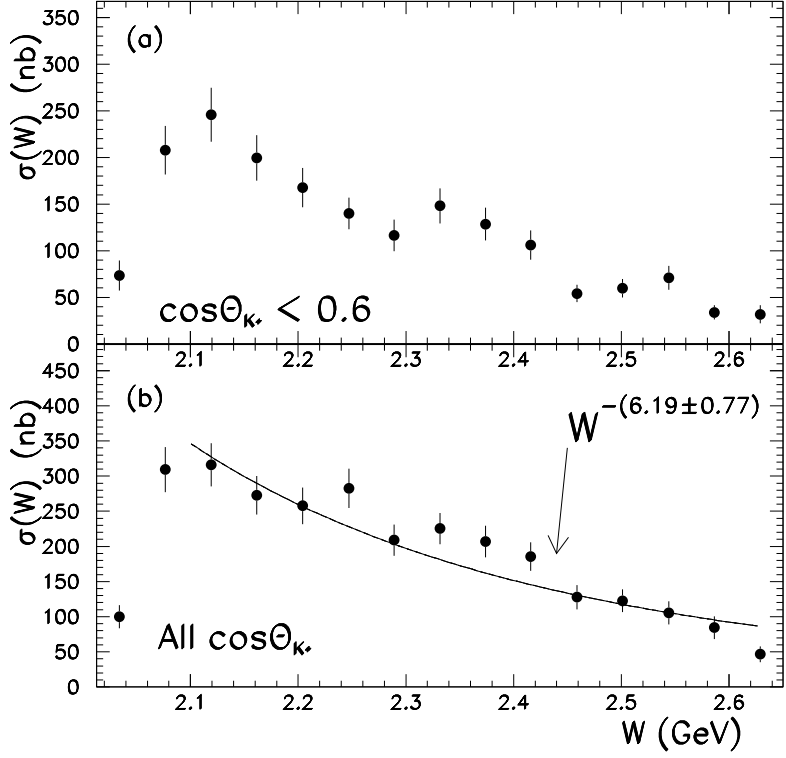}}
\caption{\small{The $\Lambda$(1520) production
cross section as a function of $W$ for (a) cos$\theta_{K^+}$ $<$ 0.6,
and (b) all cos$\theta_{K^+}$. The curve plotted in (b) is the result
of a power law fit to the $W$ dependence of the total cross section
for 2.1 $<$ $W$ $<$ 2.65 GeV.
The error bars in both plots represent statistical uncertainties only.}}
\label{fig:sig_W_all}
\end{figure}

The $W$ distributions for cos$\theta_{K^+}$ $<$ 0.6,
and all  cos$\theta_{K^+}$,  are shown in
Fig.~\ref{fig:sig_W_all}. The result of a power law fit to the
$W$ dependence of the total cross section is also shown in this figure.
Since the cos$\theta_{K^+}$ distributions
shown in Fig.~\ref{fig:w_dep_theta} are forward peaked and consistent
with {\it t}-channel dominance, the most likely kinematic regime
to observe  {\it s}-channel contributions
is for larger center-of-mass angles. There
are some structures in Fig.~\ref{fig:sig_W_all}(a) that are absent in 
Fig.~\ref{fig:sig_W_all}(b), but better statistical precision
is needed.
As was the case with a photoproduction 
measurement \cite{saphir} of the $\Lambda$(1116)
cross section, the $W$ distribution for the $\Lambda$(1520) electroproduction
process rises steeply near threshold. The cos$\theta_{K^+}$ distribution
for this region of $W$  shown in Fig.~\ref{fig:w_dep_theta} 
suggests at least two partial waves are
making significant contributions. This  is not the expected 
behavior if
this region of $W$ were dominated by a single resonance. 

\begin{figure}[htb] 
\epsfysize133mm
\centerline{\epsfbox{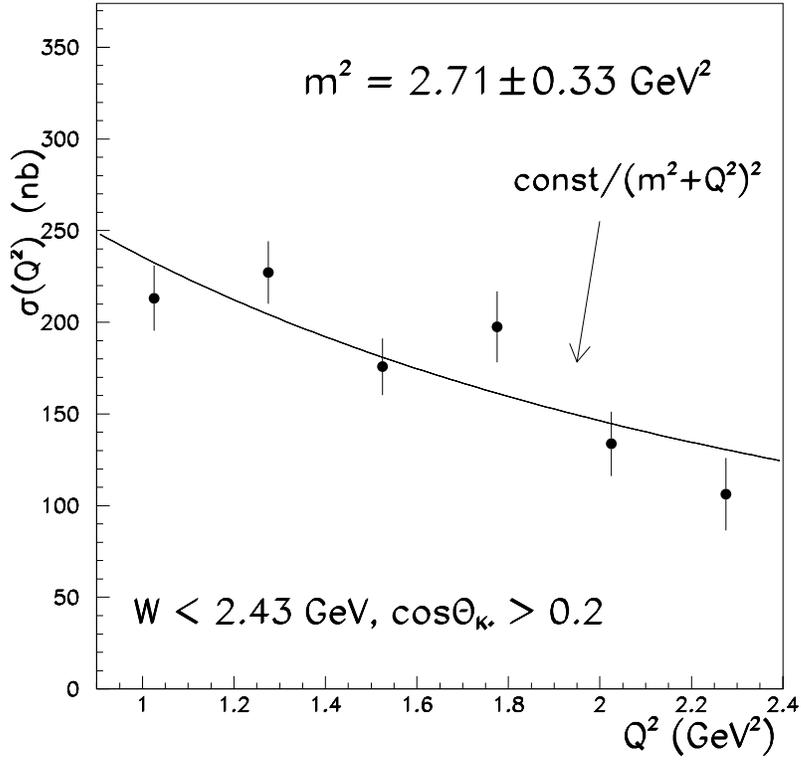}}
\caption{\small{The $Q^2$ dependence of the cross section for $W$ $<$ 2.43
GeV, and cos$\theta_{K^+}$ $>$ 0.2. The error bars represent statistical uncertainties
only. Also shown in this figure is the result of a fit to the data
of the form  $(m^2+Q^2)^{-2}$.} }
\label{fig:sig_q_dep}
\end{figure}

\begin{figure}[htb] 
\epsfysize133mm
\centerline{\epsfbox{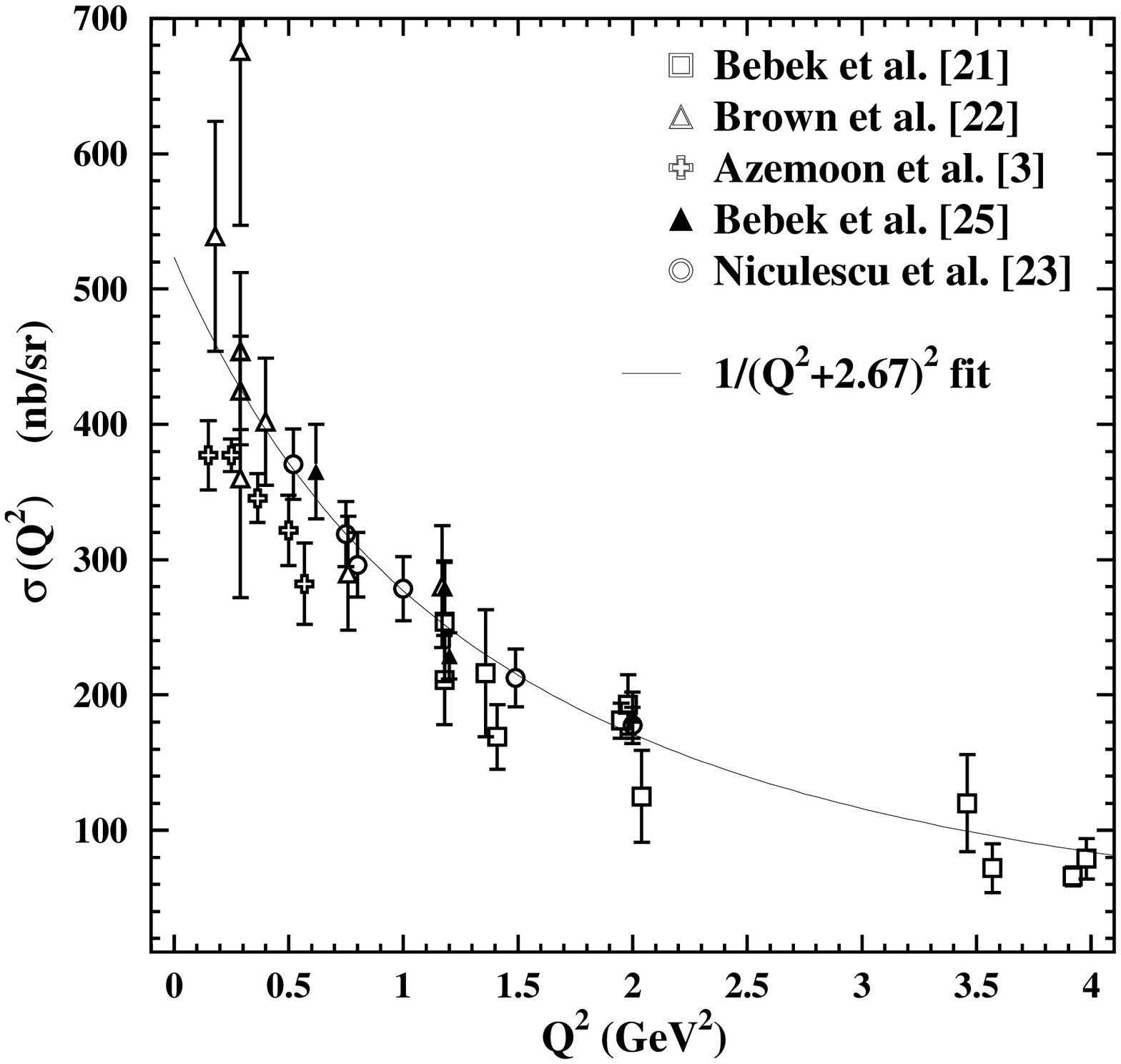}}
\caption{\small{The $Q^2$ dependence of the $\Lambda$(1116) cross section
for $W$ = 2.15 GeV.  
Some of the points have been scaled to the assumed cross section
for this $W$. Further details can be found in Ref. [21].
}}
\label{fig:q_1116}
\end{figure}

The $Q^2$ dependence of the cross section for $W$ $<$ 2.43 GeV, 
and cos$\theta_{K^+}$ $>$ 0.2, is shown in Fig.~\ref{fig:sig_q_dep}.
Previous measurements [3, 21 - 23] of the $Q^2$ dependence
of the $\Lambda$(1116) cross section studied $\theta_{K^+}$ = 0$^{\circ}$, 
and the cut on cos$\theta_{K^+}$ 
used to generate Fig.~\ref{fig:sig_q_dep}  attempts to match the 
kinematic regimes previously studied as much as possible, given
the current data set. The $Q^2$ dependence of the lighter hyperons 
is customarily parameterized assuming a $(m^2+Q^2)^{-2}$ behavior,
therefore this same function is used to parameterize the 
cross section shown in Fig.~\ref{fig:sig_q_dep}.
The fitted mass shown in
Fig.~\ref{fig:sig_q_dep} is the same (within errors) as the mass term shown in Fig.~\ref{fig:q_1116}
derived from the $\Lambda$(1116) cross 
section [3, 21 - 23] for $Q^2$ ranging from 0 to 4.0 GeV$^2$.

\begin{center}
{\large IV. Summary and discussion of results}
\end{center}

The electroproduction of the $\Lambda$(1520) strange baryon
was measured for $Q^2$ from 0.9 to 2.4 GeV$^2$, and $W$
from 1.95 to 2.65 GeV. The $\Lambda$(1520) decay angular
distributions were presented for the first time in an electroproduction
measurement, along with the cos$\theta_{K^+}$, $\phi_{K^+}$, and
{\it t}-dependencies.

Electroproduction of the $\Lambda$(1520) 
appears to be dominated by {\it t}-channel
processes, as does
the photoproduction measurement \cite{bar80}. 
The {\it t}-channel helicity frame angular distributions 
suggest longitudinal {\it t}-channel diagrams make significant
contributions to  
electroproduction but not photoproduction.
The results presented here indicate the
transition between these two sets of {\it t}-channel processes
occurs 
in the region 0 $<$ $Q^2$ $<$ 0.9 GeV$^2$, and once the transition
takes place there is little $Q^2$ dependence to the reaction
mechanism. However, it is important to keep in mind that
the ranges of some other kinematic quantities do not overlap
in these two measurements. For example, the difference between the
photo- and electroproduction spin projections could be primarily due to different $W$ 
ranges.
Most of the $\Lambda$(1520) photoproduction data of Ref. \cite{bar80} are from a  
higher region of $W$
than is presented here, and it is not unusual for the exchange of J $=$ 1 vector
mesons to make a larger contribution
for $W$ well above threshold.
The difference could  also be a consequence of different {\it t}-ranges, since Ref. \cite{bar80}
studied a range of {\it t} from $-0.2$ to $-0.65$ GeV$^2$, far above the
region of {\it t} studied in electroproduction.
The CLAS detector has recently been used to measure 
$\Lambda$(1520) photoproduction over the region
of $W$ presented here, enabling a future direct
comparison of photo- and electroproduction.
In addition, once the analysis of data taken with 3.1 and 4.8 GeV electron beam energies
is complete, the study of the spin projection ratios 
will be extended to smaller and larger values of  $Q^2$.

%Recently it has been reported \cite{nic98,moh99} that little $Q^2$ variation
%exists for the ratios of the virtual photon cross sections $\sigma_L$/$\sigma_T$ 
%for $\Lambda$(1116) production from $Q^2$ = 0.52 to 2.0 GeV$^2$.
%As shown in Fig.~\ref{fig:q_dep}, the results presented here 
%indicate little $Q^2$ dependence to the ratios of the
%$\Lambda$(1520) spin projections from $Q^2$ = 0.9 to 2.4 GeV$^2$.
%These results might be a consequence of little $Q^2$ variation to  the $\Lambda$(1520)
%virtual photon $\sigma_L$/$\sigma_T$ ratios.
%Further experimental studies of the similarities
%between $\Lambda$(1116) and $\Lambda$(1520) production 
%would be useful, in particular an L-T separation of $\Lambda$(1520)
%virtual photon cross sections.

It is interesting that  the $Q^2$ dependencies of the $\Lambda$(1520)
and $\Lambda$(1116)
cross sections both yield fitted mass values close to $m$ $\sim$ 1.65 GeV, 
while the $\Sigma$(1193)   cross section 
yields a substantially smaller value,
namely $m$ $\sim$ 0.89 GeV \cite{beb77,moh99}. 
Therefore, it might be the case that this larger mass
term is characteristic of all $\Lambda$-hyperons.
Given this possibility, it is worthwhile to revisit 
some of the original 
models, presented more than 25 years ago, that addressed
the qualitative
differences between $\Lambda$(1116) and
$\Sigma$(1193) production.

%\cite{close75,cc75}

Some of these first attempts [26-28] to explain
the difference between the $\Lambda$(1116) and
$\Sigma$(1193) cross sections assumed hyperon
production at high $Q^2$ 
is dominated by the virtual
photon scattering off one of the quarks in the proton.
The remaining two quarks couple into either isospin-zero or isospin-one 
pairings, and a few general
arguments were sufficient to show the isospin-zero 
pairing is preferred as the Bjorken $x$ variable approaches 1.0.
Therefore in these models an up quark 
interacts with the virtual photon, and the isospin-zero 
pairing of the other two quarks leads to the
preference for $\Lambda$(1116) production.
This framework also predicts the $\Sigma$(1193) cross
section drops off much more rapidly with $Q^2$ than the $\Lambda$(1116) cross section,
even for small values of $Q^2$ at which $x$ is much less than 1.0.
The fact the $Q^2$ dependencies of the $\Lambda$(1116) and
$\Lambda$(1520) cross sections are identical is consistent with
this model, and suggests the isospin of the produced hyperon
is an important quantity in determining the $Q^2$ behavior
of hyperon production.

We would like to acknowledge
the outstanding efforts of the staff of the Accelerator and Physics
Divisions at Jefferson Laboratory that made 
this measurement possible.
This work was supported by the U.S. Department of Energy and the
National Science Foundation, the French Commissariat \`{a} l'Energie
Atomique, the Italian Istituto Nazionale di Fisica Nucleare, and
the Korea Science and Engineering Foundation. We would also
like to acknowledge useful conversations
with F. Close, S. Capstick, and C. Bennhold.
The Southeastern Universities Research Association (SURA) operates the
Thomas Jefferson National Accelerator Facility for the United States
Department of Energy under contract DE-AC05-84ER40150.

\begin{center}
\begin{tabular}{|c|c|c|c|c|  }
\hline
  &  a$_{0}$     &  a$_{1}$    &  a$_{2}$   &  a$_{3}$ \\
\hline
$W$$<$2.1 & 0.60$\pm$.02  & 0.31$\pm$.06   & 0.04$\pm$.07 & 0.05$\pm$.05   \\      
2.1$<$$W$$<$2.2 & 0.59$\pm$.04  & 0.39$\pm$.04   & 0.04$\pm$.04 & $-0.03$$\pm$.03   \\      
2.2$<$$W$$<$2.3 & 0.55$\pm$.04  & 0.42$\pm$.05   & 0.05$\pm$.05 & $-0.01$$\pm$.05   \\      
2.3$<$$W$$<$2.4 & 0.49$\pm$.04  & 0.37$\pm$.05  & 0.19$\pm$.03 & $-0.06$$\pm$.04   \\      
2.4$<$$W$$<$2.5 & 0.54$\pm$.05  & 0.37$\pm$.08  & 0.20$\pm$.05 & $-0.11$$\pm$.04   \\      
2.5$<$$W$$<$2.65 & 0.40$\pm$.04  & 0.37$\pm$.06  & 0.21$\pm$.04 & 0.02$\pm$.05   \\      
\hline
\end{tabular}
\end{center}
Table 1. {\small The normalized coefficients of the Legendre polynomials
for the fits plotted in Fig.~4. The coefficients are
normalized such that they sum to unity. The uncertainty 
due to the parameterization of the background under
the $\Lambda$(1520) peak contributes an additional
uncertainty in the coefficients of $\sim$.004, which
is negligible compared to the errors shown above.}

\vspace{2 mm}

\begin{center}
\begin{tabular}{|c|c|  }
\hline
{\scriptsize Q$^2$ range (GeV$^2$)} &  {\scriptsize ratio ($m_z=\pm\frac{3}{2})/(m_z=\pm\frac{1}{2}$) }  \\
\hline
0.9-1.2 & .806$\pm$.125  \\      
1.2-1.5 & .534$\pm$.148  \\      
1.5-1.8 & .614$\pm$.108  \\      
1.8-2.4 & .558$\pm$.108  \\      
\hline
\end{tabular}
\end{center}
{\small Table 2. The ratios of the spin projection parentages
for the four regions of $Q^2$ presented in Fig.~9.}

\vspace{2 mm}

\begin{center}
\begin{tabular}{|c|c|c|c|  }
\hline
{\scriptsize Q$^2$ interval (GeV$^2$) } & {\small  A }    & {\small  B }   & {\small  C }   \\
\hline
0.9 $<$ Q$^2$ $<$ 1.2  & 1.0 $\pm$ .07  & $-0.23$ $\pm$ .10 & 0.59 $\pm$ .10  \\      
1.2 $<$ Q$^2$ $<$ 1.5 & 1.0 $\pm$ .08 & $-0.23$ $\pm$ .11 & 0.10 $\pm$ .11 \\      
1.5 $<$ Q$^2$ $<$ 1.8 & 1.0 $\pm$ .07 & $-0.16$ $\pm$ .10 & 0.27 $\pm$ .10 \\      
1.8 $<$ Q$^2$ $<$ 2.4 & 1.0 $\pm$ .07 & $-0.25$ $\pm$ .10 & 0.46 $\pm$ .09 \\      
\hline
\end{tabular}
\end{center}
Table 3. {\small Summaries of the fits shown in Fig.~13. The fits are of the form
A+B$\ast$cos2$\phi_{K^+}$+C$\ast$cos$\phi_{K^+}$, and the entries in this
table are the fitted values of the parameters with A normalized to one.}

\vspace{2 mm}

\clearpage

\clearpage

%\begin{figure}
%\epsfysize133mm
%\centerline{\epsfbox{fig_electron.ps}}
%\caption{\small{(a) The total energy deposited in the calorimeter
%(x-axis) versus the momentum of the track as derived from the
%drift chamber (y-axis) for all negatively charged tracks. (b)
%The subset of events from (a) that also recorded a hit in the Cherenkov
%detector. The vertical band of tracks in (a) that deposit a small amount of energy
%in the calorimeter, and is missing from (b), are  $\pi^-$ tracks.}}
%\label{fig:elec_id}
%\end{figure}

%\clearpage

\clearpage

\clearpage

\clearpage

\clearpage

\clearpage

\clearpage

\clearpage

\clearpage

\clearpage

\clearpage

\clearpage

\clearpage

\clearpage

\begin{thebibliography} {99}

\bibitem{bar80} D. Barber $et$ $al.$, Z. Physik C
{\bf 7}, 17 (1980).

\bibitem{boy71} A. Boyarski $et$ $al.$, Phys. Lett. B
{\bf 34}, 547 (1971).

\bibitem{aze75} T. Azemoon $et$ $al.$, Nuclear Physics
{\bf 95}, 77 (1975).

\bibitem{wjc92} R. Williams, C. Ji, and S. Cotanch, 
Phys. Rev. C {\bf 46}, 1617 (1992).

\bibitem{dav96} J. David $et$ $al.$, 
Phys. Rev. C {\bf 53}, 2613 (1996).

\bibitem{cap88} S. Capstick and W. Roberts, 
Phys. Rev. D {\bf 58}, 074011 (1998).

\bibitem{clas_nim} W. Brooks, Nucl. Phys. A 
{\bf 663-664}, 1077 (2000); B. Mecking $et$ $al.$,
in preparation.

\bibitem{dc1} D.S. Carman $et$ $al.$,
Nucl. Instr. and Methods A {\bf 419}, 315 (1998).

\bibitem{dc2} M. D. Mestayer $et$ $al.$,
Nucl. Instr. and Methods A {\bf 449}, 81 (2000).

\bibitem{sc} E. Smith $et$ $al.$, Nucl.
Instr. and Methods A {\bf 432}, 265 (1999).

\bibitem{cc} G. Adams $et$ $al.$, submitted to  
Nucl. Instr. and Methods A, (2000). 

\bibitem{ec} M. Amarian $et$ $al.$, Nucl. Instr. and Meth. A
{\bf A460} (2001) 239.
  
\bibitem{lac} M. Anghinolfi $et$ $al.$,  Nucl.
Instr. and Methods A {\bf 447}, 424 (2000).

\bibitem{trig92} D. Doughty $et$ $al.$, IEEE {\bf 18}, 241 (1992).

\bibitem{PDG}  D. Groom $et$ $al$., The European Physical Journal C 
{\bf 15}, 1 (2000). 

\bibitem{costy2k} K. Lukashin $et$ $al$., hep-ex/0101030, accepted for publication
in  Phys. Rev. C, (2001). 

\bibitem{clasnote} S. Barrow, CLAS-ANALYSIS Report 2000-002,
2000 (unpublished); CLAS-ANALYSIS Report 2001-004, 2001 (unpublished).

\bibitem{costy99} K. Loukachine, Ph.D. thesis,
Virginia Polytechnic Institute, 2000.

\bibitem{mo69} L. Mo and Y. Tsai, 
Rev. Mod. Phys. {\bf 41}, 205 (1969).

\bibitem{saphir} M. Q. Tran  $et$ $al.$, Phys.
Lett. B {\bf 445}, 20 (1998).

\bibitem{beb77} C. J. Bebek  $et$ $al.$, Phys.
Rev. D {\bf 15}, 594 (1977).

\bibitem{brown72} C. Brown   $et$ $al.$, Phys.
Rev. Lett. {\bf 28}, 1086 (1972).

\bibitem{nic98} G. Niculescu  $et$ $al.$, Phys.
Rev. Lett. {\bf 81}, 1805 (1998).

\bibitem{moh99} R. Mohring,  Ph.D. thesis,
University of Maryland, 1999.

\bibitem{beb2} C. J. Bebek  $et$ $al.$, Phys.
Rev. Lett. {\bf 32}, 21 (1974).

\bibitem{close75} F. Close, Nucl. Phys. B {\bf 73}, 410 (1974).

\bibitem{nac74} O. Nachtmann, Nucl. Phys. B {\bf 74}, 422 (1974).

\bibitem{cc75} J. Cleymans and F. Close, Nucl. Phys. B {\bf 85}, 429 (1975).

\end{thebibliography}
\end{document}